\documentclass[journal=nalefd,manuscript=article,layout=twocolumn]{achemso}

\setkeys{acs}{articletitle = true} % will suppress showing article title

\usepackage{amsmath}%
\usepackage{amsfonts}%
\usepackage{amssymb}%

\usepackage{graphicx}

\usepackage{fullpage}

\usepackage{dsfont}
\usepackage{hyperref}
\usepackage{bm}
\usepackage{multirow}
\usepackage{color}
\usepackage{float}
\usepackage{changes}

\definecolor{ablue}{rgb}{0.1,0.3,0.75}
\definecolor{agray}{rgb}{0.5,0.5,0.55}

\author{Roberto Rosati}
\affiliation{Department of Physics, Philipps-Universität Marburg, Renthof 7, D-35032 Marburg, Germany}
\email{rosatir@uni-marburg.de}

\author{Koloman Wagner}
\affiliation{Department of Physics, University of Regensburg, Regensburg D-93053, Germany}

\author{Samuel Brem}
\affiliation{Department of Physics, Philipps-Universität Marburg, Renthof 7, D-35032 Marburg, Germany}

\author{Ra\"ul Perea-Caus\'in}
\affiliation{Chalmers University of Technology, Department of Physics,
412 96 Gothenburg, Sweden}

\author{Jonas D. Ziegler}
\author{Jonas Zipfel}
\affiliation{Department of Physics, University of Regensburg, Regensburg D-93053, Germany}

\author{Takashi Taniguchi}
\affiliation{International Center for Materials Nanoarchitectonics,  National Institute for Materials Science, Tsukuba, Ibaraki 305-004, Japan}

\author{Kenji Watanabe}
\affiliation{Research Center for Functional Materials, National Institute for Materials Science, Tsukuba, Ibaraki 305-004, Japan}

\author{Alexey Chernikov}
\affiliation{Institute for Applied Physics, Dresden University of Technology, Dresden, D-01187, Germany}
\alsoaffiliation{Department of Physics, University of Regensburg, Regensburg D-93053, Germany}

\author{Ermin Malic}
\affiliation{Department of Physics, Philipps-Universität Marburg, Renthof 7, D-35032 Marburg, Germany}
\alsoaffiliation{Chalmers University of Technology, Department of Physics,
412 96 Gothenburg, Sweden}
\title{Non-equilibrium diffusion of dark excitons in atomically thin semiconductors}

\begin{document}

\begin{abstract}
 \textbf{Atomically thin semiconductors provide an excellent platform to study intriguing many-particle physics of  tightly-bound excitons. 
In particular, the properties of tungsten-based transition metal dichalcogenides are determined by a complex manifold of bright and  dark exciton states.
While dark excitons are known to  dominate the relaxation dynamics and low-temperature photoluminescence, their impact on the \textit{spatial} propagation of excitons has remained elusive.
In our joint theory-experiment study, we address this intriguing regime of dark state transport by resolving the spatio-temporal exciton dynamics in hBN-encapsulated WSe$_2$ monolayers after resonant excitation.
We find clear evidence of an unconventional, time-dependent diffusion during the first tens of picoseconds, exhibiting strong deviation from the  steady-state propagation.
Dark exciton states are initially populated by phonon emission from the bright states, resulting in  creation of hot excitons whose
rapid expansion  leads to a transient increase of the diffusion coefficient by more than one order of magnitude.
These findings are relevant for both fundamental understanding of the spatio-temporal exciton dynamics in atomically thin materials as well as their technological application by enabling rapid diffusion.
}
   \end{abstract}
\maketitle

Excitonic phenomena are known to determine the properties of atomically thin transition metal dichalcogenides (TMDs). 
A plethora of Coulomb-bound states including bright, spin- and momentum-dark excitons\cite{Wang18,Mueller18,Mak10,Xiao12,Chernikov14,Zhang15,Selig16} as well as spatially-separated interlayer excitons in van der Walls heterostructures \cite{Fogler14, Rivera2015, Ovesen19, Merkl19} dominate the optical response and the ultrafast dynamics of these technologically promising materials. 
Particularly rich excitonic manifolds are observed in tungsten-based TMDs \cite{Selig16,  Malic18, Deilmann19, Madeo20, Wallauer20, Dong20} that exhibit a multitude of sharp resonances in their photoluminescence (PL) spectra.
These resonances stem from  \textit{dark} exciton states, which can not directly interact with in-plane polarized light due to either spin- or momentum-conservation.
However, they are sufficiently long-lived and can recombine either through out-of-plane polarized or phonon-assisted emission resulting in pronounced PL signatures located energetically below the optically bright exciton  \cite{Ye18,Barbone18,Li19,Liu19,Lindlau17,He20, Brem20}.

\begin{figure}[t!]
\centering
\includegraphics[width=\linewidth]{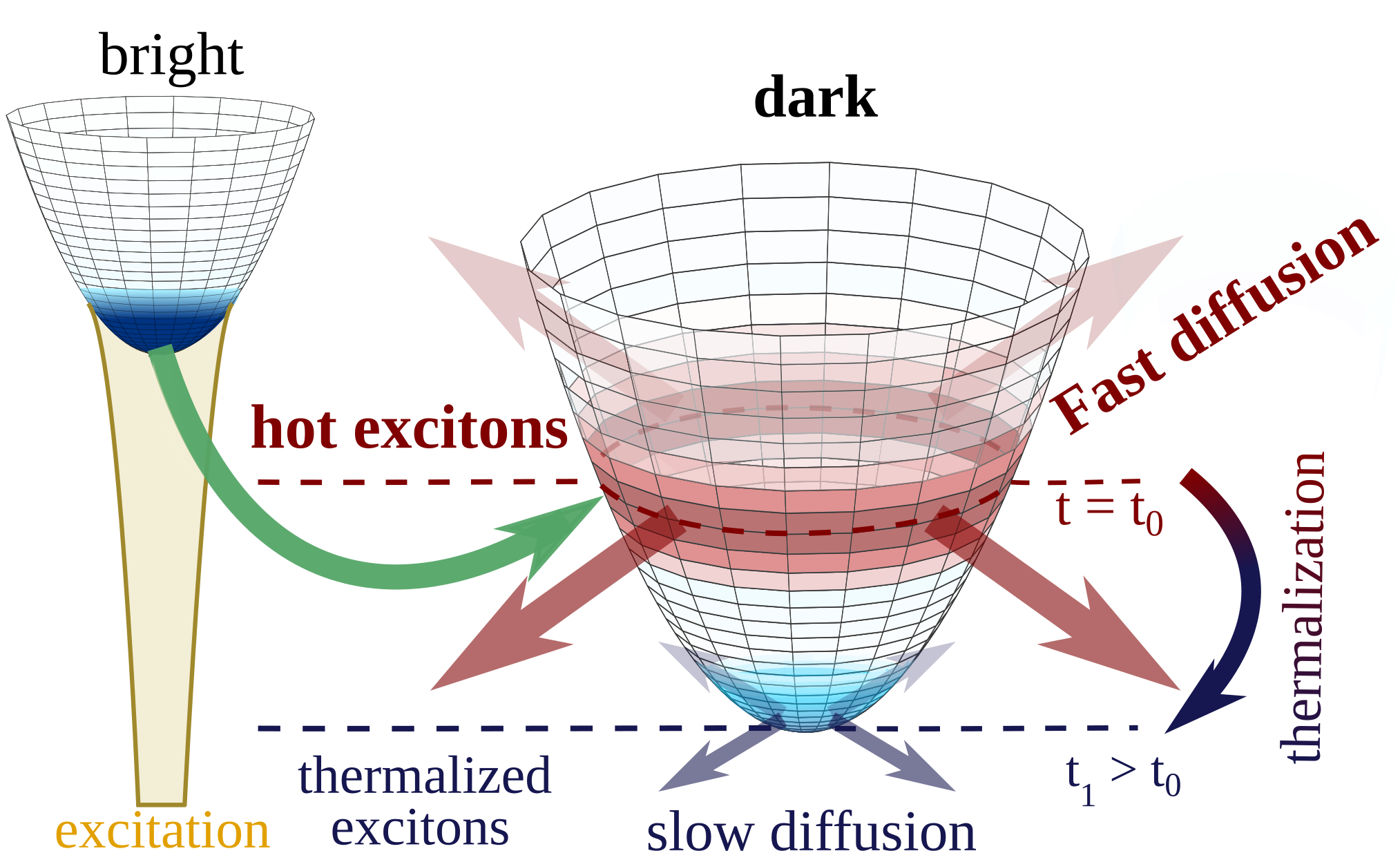}
\caption{\textbf{Hot  exciton formation.} Optically excited bright excitons scatter via emission of phonons into energetically lower momentum-dark exciton states. 
The resulting excess energy of dark excitons leads to a fast, transient diffusion of the non-equilibrium population prior to thermalization and cooling.
\label{fig:Fig1}}
\end{figure}

Following resonant optical excitation, energetically favorable dark states are populated   on a sub-picosecond timescale
via emission of phonons from the energetically higher bright states (Fig. \ref{fig:Fig1}). Due to the mismatch between exciton valley separation and phonon energy, this results in hot dark excitons exhibiting a considerable excess energy  (Fig. \ref{fig:Fig1}), as recently evidenced in spectrally and temporally resolved PL spectra \cite{Rosati20b}.
The presence of such overheated excitons in an otherwise cold lattice represents a particularly interesting scenario that should directly lead to a rapid expansion of  excitons, as schematically illustrated in Fig. \ref{fig:Fig1}.
Interestingly,  while spatio-temporal dynamics of excitons in TMD mono- and few-layer materials has attracted broad attention \cite{Kato2016,Cadiz18,Cordovilla19,Uddin20,Zipfel20,Kulig18,Cordovilla18,Wang2019,Glazov2019,Perea19,Rosati20,Harats20,Moon20,Rosati21a},
only little is known regarding transient propagation of \textit{dark} excitons.
Non-equilibrium excitons with excess energy are expected to lead to a non-conventional diffusion, as e.g. proposed for the interpretation of non-linear, room-temperature effects \cite{Cordovilla18} associated also with Auger scattering \cite{Kulig18,Perea19}.
A direct demonstration of \textit{transient} dark exciton diffusion far from the thermal equilibrium and the potential to create reasonably long-lived hot exciton populations has yet remained unexplored.

In this work, we address this intriguing exciton diffusion regime in a joint theory-experiment study, where we combine microscopic many-particle modeling with spatially and temporally resolved luminescence microscopy.
To suppress environmental disorder we employ hBN-encapsulated WSe$_2$ monolayers, cooled down to cryogenic temperatures.
These conditions allow us to take advantage of prolonged thermalization times \cite{Rosati20b} but also gain access to dark excitons through characteristic, spectrally narrow phonon sidebands \cite{Brem20,Courtade17,Ye18}.
In both theory and experiment we demonstrate an unconventional, time-dependent diffusion during thermalization and relaxation of dark excitons. 
We show a pronounced increase of the effective diffusion coefficient by more than one order of magnitude at early times after the excitation, in contrast to the stationary case of an equilibrated exciton population.
The agreement between microscopic theory and spatio-temporal PL measurements provides a clear evidence of hot dark exciton propagation in tungsten-based TMD monolayers. 

\textbf{Microscopic modeling:} 
First, we present theoretical calculations with the goal to microscopically describe spatially and temporally resolved dynamics of the phonon-assisted emission from dark excitons.
To obtain excitonic energies and wavefunctions we solve the Wannier equation \cite{Haug09,Berkelbach2015,Selig16} including material-specific parameters for the electronic bandstructure \cite{Kormanyos15}.
Due to the resonant optical excitation and considerable binding energies, we focus our study on the 
energetically lowest $1s$ states of bright KK and momentum-dark K$\Lambda$ and KK$^\prime$ excitons.
Here, K, K$^\prime$, $\Lambda$ denote the high-symmetry points in the reciprocal space that host electrons and holes forming an exciton.
We restrict our analysis on processes that do not require a spin-flip.  Spin-dark states are expected to have a minor quantitative impact on the considered non-equilibrium phenomena due to their much slower thermalization \cite{Song13} and predominant relevance of spin-conserving processes in the exciton relaxation.
In close agreement with  first-principles studies \cite{Deilmann19}, we find that in WSe$_2$ monolayers KK$^\prime$ excitons are the energetically lowest states followed by K$\Lambda$ and KK excitons. 

To obtain a microscopic access to the spatio-temporal exciton dynamics, we introduce the excitonic Wigner function $N^v_{\mathbf{Q}}(\mathbf{r},t)$, which provides the quasi-probability of finding  excitons with momentum $\mathbf{Q}$ at time $t$ in position $\mathbf{r}$ and exciton valley $v$ \cite{Rosati20}. 
We derive its equation of motion using the Heisenberg equation and applying the transformation into the Wigner representation \cite{Hess96}. 
In the low-density excitation regime it reads\cite{Rosati20}
\begin{align}
\label{SBE}
\nonumber
\dot{N}^v_\mathbf{Q}(\mathbf{r},t)&\hspace{-2pt}= \hspace{-2pt}\left(\frac{\hbar \mathbf{Q}}{M_v}\cdot \nabla - \gamma \delta_{\mathbf{Q},0}\delta_{v,\text{KK}} \right)\hspace{-2pt}N^v_\mathbf{Q}(\mathbf{r},t)\\ 
&+\Gamma^{v;\text{KK}}_{\mathbf{Q};0} |p_{0}(\mathbf{r},t)|^2\!\!+\!\!\left.\dot{N}^v_\mathbf{Q}(\mathbf{r},t)\right|_{th}\hspace{-5pt}.
\end{align}
The first term describes free propagation of excitons depending on the total mass $M_v$ and the spatial gradient in the exciton occupation.
The second term takes into account losses due to the radiative recombination $\gamma$ within the light cone ($\delta_{\mathbf{Q},0}\delta_{v,KK}$) \cite{Selig18}. 
The third and fourth contributions represent phonon-assisted formation and thermalization of excitons, respectively.  

For our study, we consider a short and confined optical pulse that is tuned into resonance with the bright exciton states and initially creates an excitonic polarization P$_{\mathbf{Q}\approx 0}(\mathbf{r}, t)$ in the light cone (often denoted as ``coherent excitons'' in the literature \cite{Kira06}). 
Exciton-phonon scattering with the rates $\Gamma^{vv^\prime}_{\mathbf{Q}\mathbf{Q}^\prime}$ drives the formation of incoherent excitons also outside of the light cone. 
Finally, exciton thermalization is described via the Boltzmann scattering term  $\left.\dot{N}^v_{\mathbf{Q}}(\mathbf{r},t)\right|_{th}=\Gamma^{\text{in},v}_{\mathbf{Q}}(\mathbf{r},t)\!-\!\Gamma^{\text{out},v}_{\mathbf{Q}}N_{\mathbf{Q}}(\mathbf{r},t)$ with phonon-driven in- and out-scattering rates $\Gamma^{\text{in/out},v}_{\mathbf{Q}}(\mathbf{r},t)$\cite{Selig18, Brem18}.

The first contribution in Eq. (\ref{SBE}) alone leads to a ballistic exciton propagation, where each state moves in space along the direction of $\mathbf{Q}$ with a velocity proportional to $\vert\mathbf{Q}\vert$. 
Exciton-phonon scattering (second line in Eq. (\ref{SBE})), however, redistributes exciton occupation primarily toward states with smaller energies (energy-relaxation) and those of different orientation in the reciprocal space (momentum-relaxation). 
In contrast to the ballistic regime, these scattering processes lead to a typically slower, diffusive exciton propagation. 

Different propagation regimes can be quantified by studying the evolution of the spatial broadening $w(t)$ of a given exciton distribution $N(\mathbf{r},t)$. This is proportional to the standard deviation, i.e. $w^2(r)=\int \mathbf{r}^2N(\mathbf{r},t) d\mathbf{r}/\int N(\mathbf{r},t) d\mathbf{r}$, and corresponds for a Gaussian distribution to the denominator in the exponent ( $N(\mathbf{r},t)\propto\exp{[-x^2/w^2(t)]}$). 
In the ballistic regime, $w^2$ would increase quadratically over time, while in the conventional diffusive regime the dependence is strictly linear \cite{Steininger97}. 
Deviations from this conventional linear law can be described by defining an effective \textit{time-dependent} diffusion coefficient $D(t)=\frac{1}{4}\partial_t w^2$ \cite{Rosati20}.
This coefficient typically converges to a constant $D=k_B T \tau_s/M_X$, when the stationary diffusion is reached, with $\tau_s$ being a state-independent scattering time. 
As we demonstrate in the following, the regime of unconventional, time-dependent diffusion can last as long as several tens of ps  in WSe$_2$ monolayers at cryogenic temperatures, where equilibration mechanisms are sufficiently slow. 

\begin{figure}[t!]
\centering
\includegraphics[width=\linewidth]{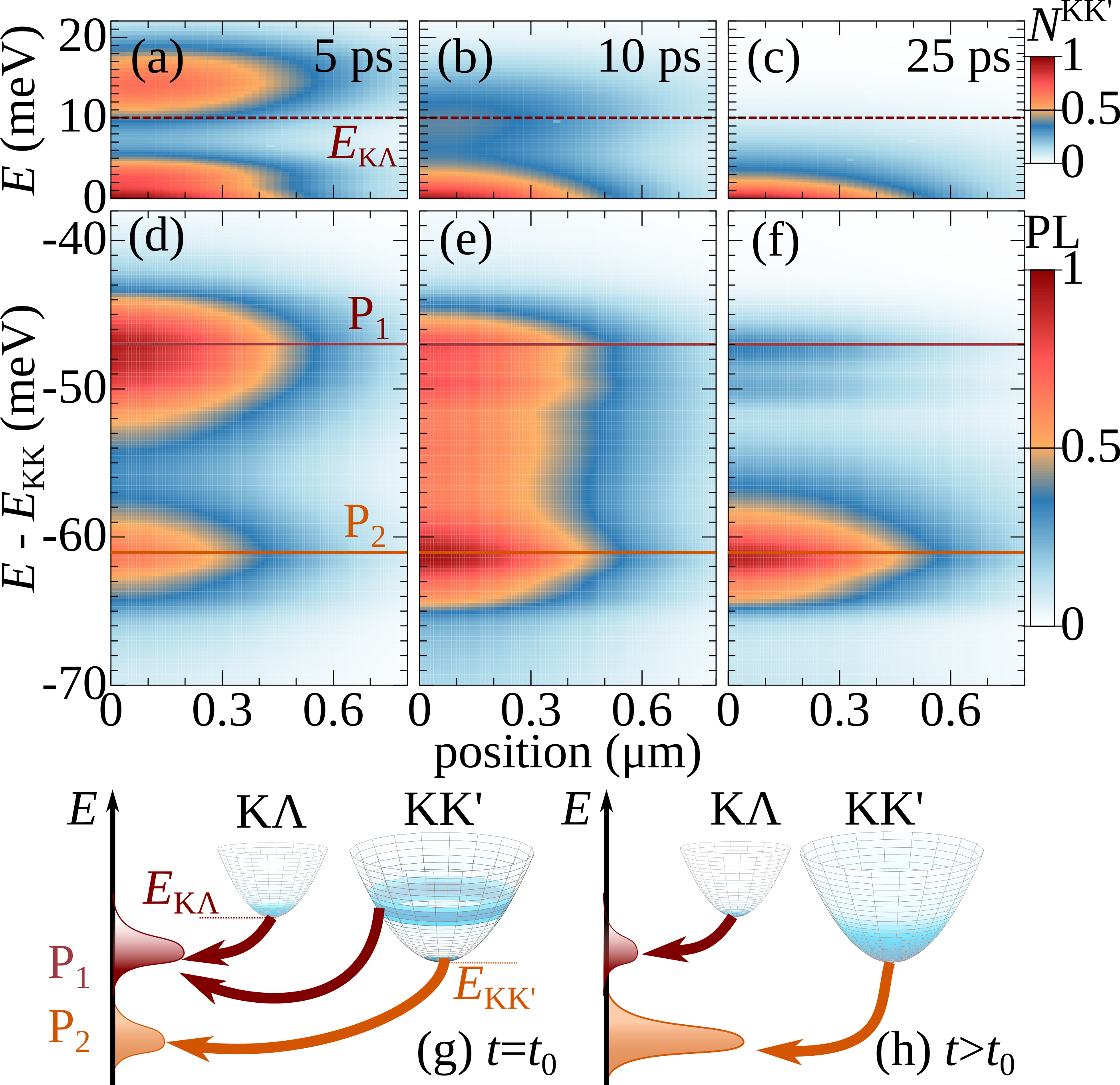}
\setlength{\belowcaptionskip}{-10pt}
\caption{\textbf{Non-equilibrium excitons dynamics.} 
 (a)-(c) Spatio-temporal dynamics of momentum-dark KK$^\prime$ excitons at 20 K illustrating the interplay between exciton propagation away from the excitation spot and their thermalization towards a stationary distribution. 
(d)-(f) Corresponding spatio-temporal evolution of the PL from momentum-dark excitons emitting through phonon-assisted recombination. The two phonon sidebands P$_1$ and P$_2$ can be traced back mainly to  KK$^\prime$ excitons with some contribution  from K$\Lambda$ states at early times. 
 Schematic illustration of the origin of  phonon sidebands (g) directly after optical excitation and (h) after thermalization into an equilibrium distribution.
\label{fig:Fig2}}
\end{figure}

\textbf{Spatio-temporal exciton dynamics:}
To investigate the spatially and temporally dependent optical response of hBN-encapsulated WSe$_2$ monolayers, we solve the equation of motion for the Wigner function [Eq. (\ref{SBE})] and the generalized Elliot formula for the PL (cf. SI).
In the calculations, we set the lattice temperature to 20 K and consider pulsed, confined resonant excitation of the bright KK exciton state with an initial spatial  width of $w\approx0.5\mu$m. 
Figures \ref{fig:Fig2}(a-c) illustrate the resulting occupation of the energetically lowest momentum-dark KK$^\prime$ excitons, $N^{\text{KK}^\prime}_{\mathbf{Q}}(\mathbf{r},t)$ at different times after the optical excitation.
It is sufficient to focus on only one spatial direction due to rotational symmetry of the system. 
Overall, the exciton occupation broadens with time as excitons diffuse away from the center of the excitation spot. Their distribution in energy, however, is notably different between the three considered times of 5, 15, and 25 ps after the excitation. 
In particular, after 5 ps (Fig. \ref{fig:Fig2}(a)) excitons are still far from the thermal equilibrium. We find a pronounced hot-exciton region, where excitons carry a considerable excess energy on the order of 15 meV.
This energy is acquired after phonon-assisted relaxation from the bright to the dark state, (Fig. \ref{fig:Fig1}) and stems from the difference between the bright-dark energy separation and the energy of the inter-valley phonons involved.  
These overheated excitons subsequently thermalize and lose their kinetic energy mainly via scattering with intravalley acoustic modes, so that the distribution starts spreading approximately 10 ps after the excitation. 
After 20 to 30 ps, excitons finally form a Boltzmann distribution with a temperature corresponding to that of the lattice [Fig. \ref{fig:Fig2}(c)]. 
The presented relaxation dynamics in momentum space is  consistent with the one for the case of a spatially-\textit{homogeneous} excitation \cite{Rosati20b}.
For spatially-\textit{localized} excitation, however, it has major implications for the exciton propagation, as discussed below. 

To obtain a key observable accessible in  experiments, we present the corresponding spatially and spectrally dependent PL in Figs. \ref{fig:Fig2}(d-f).
In the studied regime, it is dominated by phonon sidebands of momentum-dark excitons, located approximately 50 to 60 meV below the energy of the bright KK exciton \cite{Brem20}.
The majority of the PL signal is traced back to the KK$^\prime$ excitons that recombine under emission of the zone-edge acoustic phonons.
Notably, phonon-assisted processes allow for all exciton states to emit, regardless of their crystal and center-of-mass momenta.
As a consequence, the profiles in Figs.\ref{fig:Fig2}(d-f) largely follow the exciton distribution presented in Figs.\ref{fig:Fig2}(a-c).
In addition to KK$^\prime$ sidebands, we also find
contributions from K$\Lambda$ excitons at early times (Figs.\ref{fig:Fig2}(g)-(h)) that partially overlap the emission from hot KK$^\prime$ excitons with an excess energy of about 15 meV (Figs.\ref{fig:Fig2}(a)). This  overlap results in an initially more pronounced P$_1$ sideband, cf. Figs.\ref{fig:Fig2}(d).

\begin{figure}[t!]
\centering
\includegraphics[width=\linewidth]{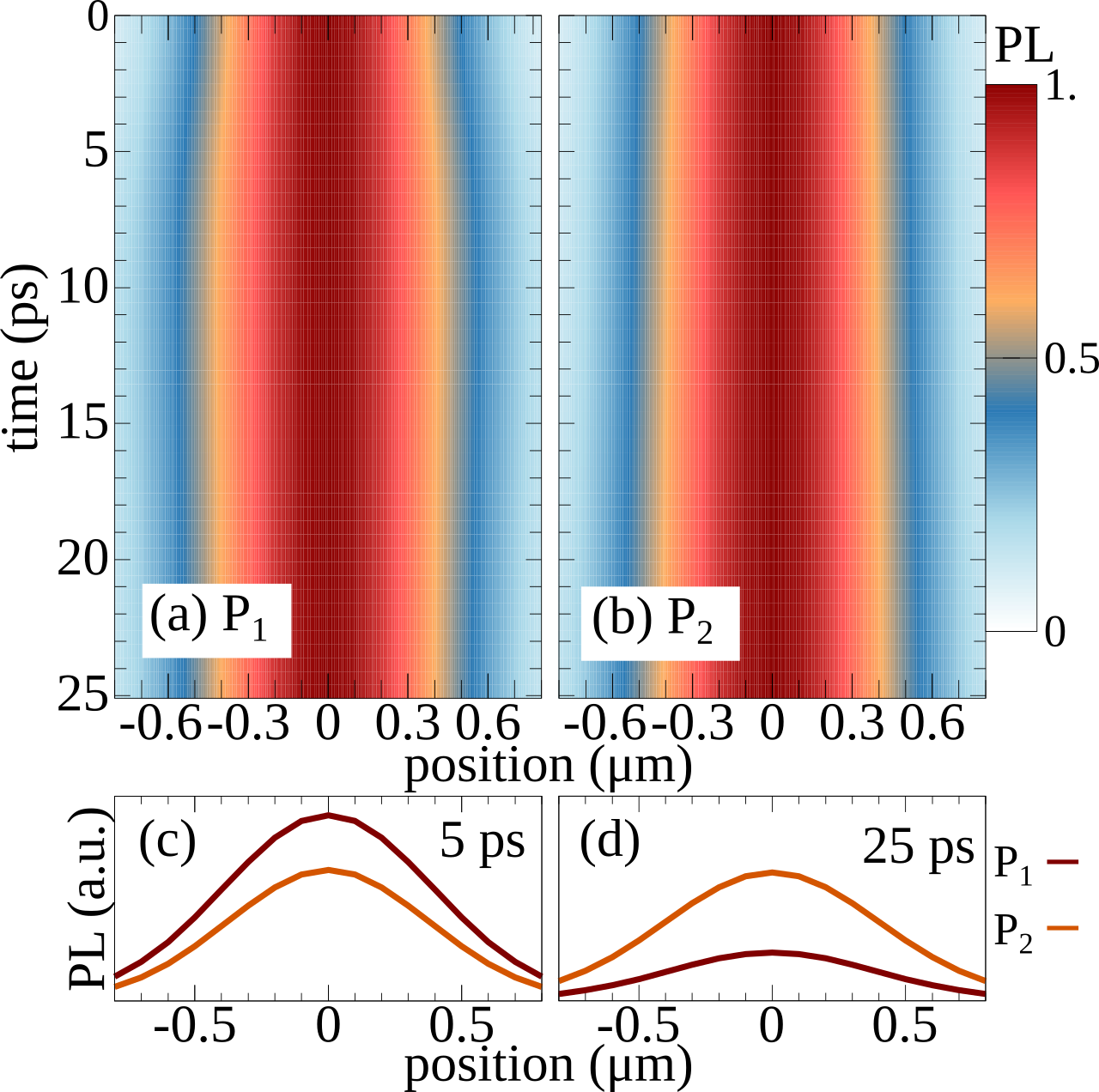}
\setlength{\belowcaptionskip}{-10pt}
\caption{\textbf{Spatio-temporal dynamics of phonon sidebands.} (a) Spectrally-resolved spatial broadening of the  P$_1$ phonon sideband shows an initial speed-up due to hot KK$^\prime$ excitons, while (b) P$_2$ exhibits a slower and prolonged exciton diffusion. 
(c)-(d) Direct comparison between the spatial intensity of P$_1$ and P$_2$ phonon sidebands at different times. The PL is normalized to the intensity of  the P$_2$ signal.  
\label{fig:Fig3}}
\end{figure}

\textbf{Transient exciton diffusion:}
Now, we consider the consequences of hot dark excitons with high excess energies for the  exciton diffusion. 
Spatio-temporal PL signals, evaluated at the energies of the two pronounced P$_1$ and P$_2$ phonon sidebands are presented in Figs. \ref{fig:Fig3}(a) and (b), respectively.
We find a fast spatial broadening of the P$_1$ signal in the first 10 ps (reaching a transient diffusion coefficient $D$ of almost 35 cm$^2$/s, cf. SI) followed by a much slower diffusion. The driving force for the initial increased spatial broadening are  hot KK$^\prime$ excitons, which emit light approximately at the same energy as the K$\Lambda$ state (Fig. \ref{fig:Fig2}(g)). This is further confirmed by calculating  energy- and exciton-valley-specific diffusion coefficients $D_{E,v}$ (cf. SI).

The situation in  P$_2$ is qualitatively different exhibiting no initial speed-up of the spatial broadening (Fig. \ref{fig:Fig3}(b)). Here, the main contribution of the PL signal stems from nearly thermalized KK$^\prime$ excitons with a vanishing excess energy (Fig. \ref{fig:Fig2}(g)). 
As a consequence, the initial spatial broadening is considerably slower compared to P$_1$, however, it lasts longer due to the thermalizing hot KK$^\prime$ excitons, cf. SI. After a few tens of ps, the stationary situation of conventional diffusion is recovered. %, where the overall diffusion coefficient reaches a value of almost 1.9 cm$^2$/s. 
  Besides the qualitatively different diffusion behavior, P$_1$ and P$_2$ phonon sidebands also differ in their intensity.  While directly after the  optical excitation P$_1$ is clearly the most pronounced PL signal [Fig. \ref{fig:Fig3}(c)], P$_2$ becomes  dominant after approx. 10 ps [Fig. \ref{fig:Fig3}(d)]. This reflects the vanishing contribution of hot KK$^\prime$ excitons to the P$_1$ signal as a thermalized distribution is approached (Fig. \ref{fig:Fig2}(h)).

\textbf{Measurement of transient exciton diffusion:}
To study the theoretically predicted impact of hot dark excitons on the transient diffusion, we perform measurements of spatially and temporally resolved PL on hBN-encapsulated WSe$_2$ monolayers, cooled to liquid helium temperature. The samples are obtained by mechanical exfoliation and stamping of bulk crystals onto SiO$_2$/Si substrates \cite{Castellanos-Gomez2014a} and allow for an effective suppression of the environmental disorder and thus offer clean access to the phonon-assisted emission from dark states.
For excitation, we use a 80\,Mhz, 100\,fs-pulsed Ti:sapphire source tuned into resonance conditions with the bright exciton $X_0$ at 1.726\,eV and focused the light to a spot with a sub-micron diameter.
The resulting emission is collected from a lateral cross-section and guided through an imaging spectrometer equipped with a mirror and a grating to provide spatial and spectral resolutions, respectively.
For time-resolved detection we use a streak camera operated in the single-photon-counting mode \cite{Kulig18}.

\begin{figure}[t!]
\centering
\includegraphics[width=\linewidth]{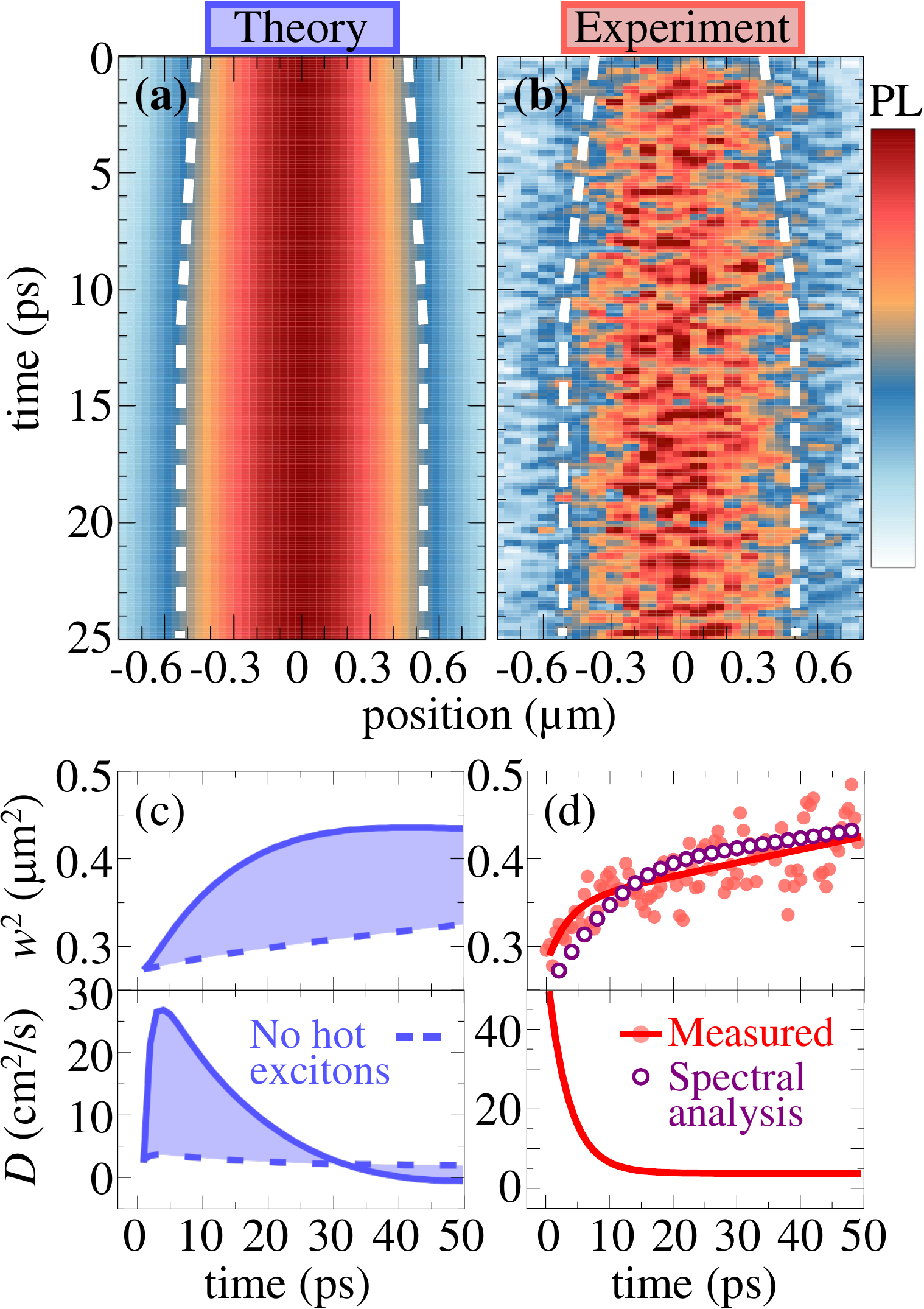}
\setlength{\belowcaptionskip}{-10pt}
\caption{\textbf{Theory-experiment comparison of hot exciton diffusion.}
 (a)-(b) Spectrally-integrated low-temperature PL illustrating the accelerated spatial broadening during the first 10 ps stemming from excitons with large excess energies. 
 (c)-(d) Temporal evolution of the squared spatial width $w^2$ and the resulting effective diffusion coefficient $D$. 
The shaded area between full evolution (solid thin line) and  assuming initial thermal equilibrium (dashed line) directly reflects the impact of non-equilibrated, hot excitons.
\label{fig:Fig4}}
\end{figure}

The spatially-resolved PL is acquired in the spectral region of the phonon sidebands including P$_1$ and P$_2$, i.e. approximately 50 meV below the bright exciton resonance, cf. the corresponding PL spectra  in  SI.   
A direct comparison between theoretically predicted and experimentally measured spatio-temporal PL is presented in Figs. \ref{fig:Fig4}(a) and (b).
The corresponding transient broadening of the spatial profiles, $w^2$, is shown in Figs. \ref{fig:Fig4}(c) and (d), including the extracted, time-dependent effective diffusion coefficient $D(t)=\frac{1}{4}\partial_t w^2$.
The time-dependent spatial expansion of  dark excitons strongly deviates from the standard diffusion law $w^2 \propto t$, exhibiting unconventional behavior during the first few tens of ps.
We observe an initially fast increase from rapidly diffusing hot excitons with high excess energies  converging towards steady-state diffusion after exciton thermalization and cooling.
The corresponding effective diffusion coefficients are found experimentally to be as high as  50 cm$^2$/s immediately after optical excitation, which is more than an order of magnitude higher than the equilibrated value of about 3.6 cm$^2$/s.
Overall, the experimental findings are in an excellent qualitative agreement with theoretical predictions.
The initially rapid diffusion from hot excitons and a substantially slower propagation in the steady-state, inherently limited by the exciton scattering with linear acoustic phonons, are captured both in experiment and theory.

In addition, it is further instructive to consider a direct comparison of spectrally and spatially resolved data.
In particular, we estimate transient diffusion coefficients that are expected from the experimentally measured spectral shifts $\Delta E(t)$ of the phonon sideband emission.
This value is obtained from the time-dependent center-of-mass of the PL spectrum in the region of P$_1$ and P$_2$ resonances (cf. SI). It should thus approximately correspond to the excess energy which enters the expression for the diffusion coefficient $D(t) = (\Delta E(t)+k_B T) \times \tau / M_X$ with T set to the lattice temperature of 5\,K,  the scattering time $\tau$ set to match the steady state value of 2.8 ps and $M_X$ fixed to 0.75$m_0$ for the dark states in WSe$_2$ \cite{Kormanyos15}. 
As shown by the circles in Fig. \ref{fig:Fig4}(d), this results in a very  good agreement with the direct measurement of the time-dependent diffusion, further supporting the interpretation of hot exciton propagation.

The diffusion behavior is found to be drastically different when neglecting the impact of hot excitons, i.e. considering an initially thermalized exciton distribution in our calculations, cf. the dashed lines in Fig. \ref{fig:Fig4}(c).
In that case, we still find a small initial increase in diffusion, however the maximum value of $D(t) < $4 cm$^2$/s is one order of magnitude smaller compared to the case of hot excitons. 
The origin of this residual effect is an initially ballistic diffusion, which drives the spatial broadening before being counteracted by phonon-assisted momentum relaxation. 
However, there is no energy relaxation involved, as excitons are already thermalized in their equilibrium ground-state. 

Interestingly, the average exciton-phonon scattering time on the order of several picoseconds also implies that during the initially fast propagation of hot excitons there are only very few exciton-phonon scattering events.
Thus, the range of the first 10 ps after the excitation represents an intermediate regime between ballistic and diffusive exciton propagation.
Furthermore, at approximately 40 ps after the optical excitation, we also predict small negative diffusion values (\ref{fig:Fig4}(c)) induced by the interstate thermalization and back-scattering processes \cite{Rosati20}. 
This weak feature could not be resolved in the experiment at the current stage, motivating future studies. 
 
In summary, we have demonstrated non-equilibrium propagation of hot dark excitons in atomically thin semiconductors by combining microscopic many-particle theory with low-temperature transient PL microscopy.
We find a fast and unconventional, time-dependent exciton diffusion with an initial increase of the diffusion coefficient of up to 50 cm$^2/s$ stemming from hot dark excitons with substantial excess energies.
This rapid expansion is followed by thermalization and cooling of the exciton distribution on a timescale  of a few tens of picoseconds, converging towards a steady-state diffusivity of about 3.6 cm$^2$/s. Our findings provide fundamental insights into the non-equilibrium transport of excitons in atomically thin semiconductors with implications towards their technological application in optoelectronic devices.%

\textbf{Supporting Information}

Additional details on theoretical and experimental methods  and analysis.
%as well as on energy-resolved PL evolution in space and time are included, discussing in particular the signal from different exciton valleys, spatial positions and initial localizations. Furthermore, a direct comparison of experimentally measured effective diffusion coefficients and excess energy/temperature are provided. 

\textbf{Acknowledgments:}
We thank Mikhail M. Glazov (Ioffe Institute) for valuable discussions on exciton diffusion.
This project has received funding from  the European
Union’s Horizon 2020 research and innovation programme
under grant agreement no. 881603 (Graphene Flagship) and from the 2D Tech VINNOVA competence Center (Ref. 2019-00068). 
The computations were enabled by resources provided by the Swedish National Infrastructure for Computing (SNIC).
Financial support by the DFG via SFB 1083 (project B9) and SFB 1244 (project B05), SPP 2196 Priority Program (CH 1672/3-1) and Emmy Noether Initiative (CH 1672/1) is gratefully acknowledged.
K.W. and T.T. acknowledge support from the Elemental Strategy Initiative
conducted by the MEXT, Japan, Grant Number JPMXP0112101001,  JSPS
KAKENHI Grant Numbers JP20H00354 and the CREST(JPMJCR15F3), JST.

\bibliography{rosatiBib_short,references}

%\begin{thebibliography}{30}

%\bibitem{Rosati20c}	
%Rosati, R., Wagner,  Brem S.,Perea-Caus\'in, R., Wietek, E., Zipfel, J.,  Ziegler, J.D., Selig, M., Taniguchi, T.,  Watanabe, K., Knorr, A., Chernikov, A. and Malic, E.  {\em ACS Photonics} \textbf{7}, 2756  
% (2020).

\newpage

\begin{tocentry}
Combining microscopic many-particle theory and low-temperature spatio-temporal photoluminescence experiments we reveal an unconventional, time-dependent exciton diffusion in  atomically thin semiconductors. 
This behavior is shown to originate from hot dark excitons with large excess energies.

\end{tocentry}

\begin{figure}[t!]
\centering
\includegraphics[width=\linewidth]{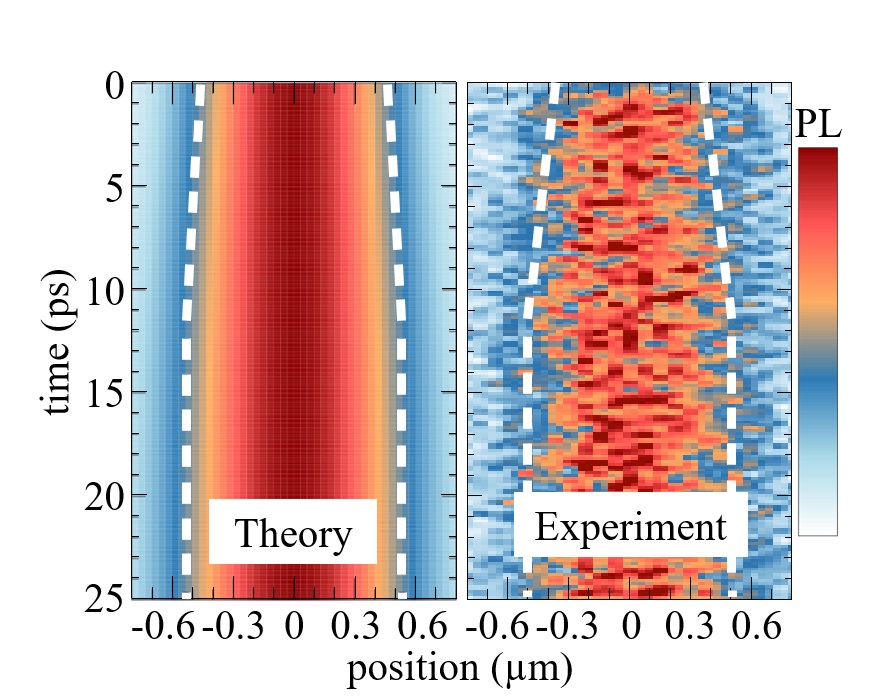}
\caption{For Table of Contents Only}
\end{figure}

%\end{thebibliography}
\end{document}

\title{Supplemental Material:\\ Non-equilibrium diﬀusion of dark excitons in atomically thin semiconductors}

\author{Roberto Rosati}
\affiliation{Department of Physics, Philipps-Universität Marburg, Renthof 7, D-35032 Marburg, Germany}
%\author{Ra\"ul Perea-Caus\'in}

\author{Koloman Wagner}
\affiliation{Department of Physics, University of Regensburg, Regensburg D-93053, Germany}

\author{Samuel Brem}
\affiliation{Department of Physics, Philipps-Universität Marburg, Renthof 7, D-35032 Marburg, Germany}

\author{Ra\"ul Perea-Caus\'in}
\affiliation{Chalmers University of Technology, Department of Physics,
412 96 Gothenburg, Sweden}

% \author{Edith Wietek}
\author{Jonas D. Ziegler}
\author{Jonas Zipfel}
\affiliation{Department of Physics, University of Regensburg, Regensburg D-93053, Germany}

\author{Takashi Taniguchi}
\affiliation{International Center for Materials Nanoarchitectonics,  National Institute for Materials Science, Tsukuba, Ibaraki 305-004, Japan}

\author{Kenji Watanabe}
\affiliation{Research Center for Functional Materials, National Institute for Materials Science, Tsukuba, Ibaraki 305-004, Japan}

\author{Alexey Chernikov}
\affiliation{Department of Physics, University of Regensburg, Regensburg D-93053, Germany}

\author{Ermin Malic}

\affiliation{Department of Physics, Philipps-Universität Marburg, Renthof 7, D-35032 Marburg, Germany}
\affiliation{Chalmers University of Technology, Department of Physics,
412 96 Gothenburg, Sweden}

\maketitle

\section{Theoretical Methods}

The excitonic landscape is calculated microscopically by solving the  Wannier equation \cite{Selig16, Brem18}
\begin{equation}
	\frac{\hbar^{2}k^{2}}{2m_{v}}\Psi_{v}(\mathbf{k})-\sum_{\mathbf{q}}W_{\mathbf{q}}\Psi_{v}(\mathbf{k}+\mathbf{q})=E_{v}^{\textrm{b}}\Psi_{v}(\mathbf{k})\quad,\label{Wan}
\end{equation}
where  $E_{v}^{\textrm{b}}$ is the exciton binding energy and $m_{v}$ the reduced exciton mass in the exciton valley. Single-particle energies and effective masses  are obtained  from first-principles calculations \cite{Kormanyos15}.  Furthermore,  $\Psi_{v}(\mathbf{k})$ describes the excitonic wave function in momentum space,
while $W_{\mathbf{q}}=V_{q}/\epsilon_{scr}(q)$ is the Coulomb potential for charges in a thin film of
thickness $d$ surrounded by a dielectric environment. It is determined by  the bare 2D-Fourier transformed
Coulomb potential $V_{q}$ and a non-local screening \cite{Brem19b} \begin{equation}
    \epsilon_{scr}(q)=\kappa_{1}\tanh\left(\frac{1}{2}\left[\alpha_{1}dq-\ln\left(\frac{\kappa_{1}-\kappa_{2}}{\kappa_{1}+\kappa_{2}}\right)\right]\right),
\end{equation}
where $\kappa_{i}=\sqrt{\epsilon_{i}^{\parallel}\epsilon_{i}^{\bot}}$
and $\alpha_{i}=\sqrt{\epsilon_{i}^{\parallel}/\epsilon_{i}^{\bot}}$
account for the parallel and perpendicular component of the dielectric
tensor $\epsilon_{i}$ of monolayer (i = 1) \cite{Laturia18} and environment (i = 2) \cite{Geick66}.  The solution of Eq. (\ref{Wan}) provides us with a set of excitonic states and associated energies. Based on these, we can study the excitonic transport via Wigner representation \cite{Hess96,Jago19}, providing in particular Eq. (1) of the main manuscript. Note that the latter holds in the low excitation regime, while at higher exciton densities Auger scattering and associated local heating can take place, resulting in the formation of halos \cite{Kulig18,Perea19,Zipfel20}.  Due to large spectral separations, we focus on the energetically lowest 1$s$ states of the three most relevant exciton valleys (KK, KK$^\prime$ and K$\Lambda$).  Furthermore, we focus on the PL stemming from momentum-dark excitons, while  spin-dark states require spin-flip processes and are thus expected to have a minor influence on the considered  spatiotemporal exciton dynamics.

 Extending the generalized Elliott formula introduced in \cite{Brem20} and adapted for time-resolved photoluminescence (PL) in \cite{Rosati20b}, we can determine  the spatiotemporally-resolved PL
  \begin{equation}
     I(E,\mathbf{r},t)=\frac{8|M|^2}{\hbar}\frac{I_0(E,\mathbf{r},t) + \sum_vI_v(E,\mathbf{r},t)}{4\left(E-E_{\text{KK}}\right)^2+\left(\gamma+\Gamma^{\text{out},\text{KK}}_{0}\right)^2},
 \end{equation}
 where $I_0(E,\mathbf{r},t)=\gamma N^{\text{KK}}_{0}/2$ provides the direct radiative recombination term, while $$I_v(E,\mathbf{r},t)=\sum_{\mathbf{Q},v,\beta,\pm} |D^v_{\beta;\mathbf{Q}}|^2\eta_{\beta,\pm} N^v_{\mathbf{Q}}(E,\mathbf{r},t)\frac{2\Gamma^{\text{out},v}_{\mathbf{Q}}}{4\left(E^v_{\mathbf{Q}}\pm \epsilon_\beta - E\right)^2 + \left(\Gamma^{\text{out},v}_{\mathbf{Q}}\right)^2}$$ provides the indirect phonon-assisted PL. Here, $M$ describes the excitonic optical matrix elements \cite{Selig18,Brem18}, the indices $\pm$ and $\beta$ refer to the emission/absorption of a phonon in the mode $\beta$ and energy $\epsilon_\beta$ and inducing an exciton-phonon coupling $D^v_{\beta;\mathbf{Q}}$.
Based on  $I_v(E,\mathbf{r},t)$ one can define a set of effective diffusion coefficients $D_{E,v}=\frac{1}{2}\partial_t \sigma_{E,v}^2$, where $\sigma_{E,v}^2(t)=\int \mathbf{r}^2I_v(E,\mathbf{r},t) d\mathbf{r}/2\int I_v(E,\mathbf{r},t) d\mathbf{r}$ is the squared  standard deviation of the spatiotemporal distribution $I_v(E,\mathbf{r},t)$ for given $v$ and $E$. Analogously one can define  $D_{E}=\frac{1}{2}\partial_t \sigma_{E}^2$, where $\sigma_{E}^2$ is the spatial width of the total PL.

\section{Valley- and energy-resolved effective diffusion coefficients}\label{sec:DE}

\begin{figure}[t!]
	\centering
			\includegraphics[width=16 cm]{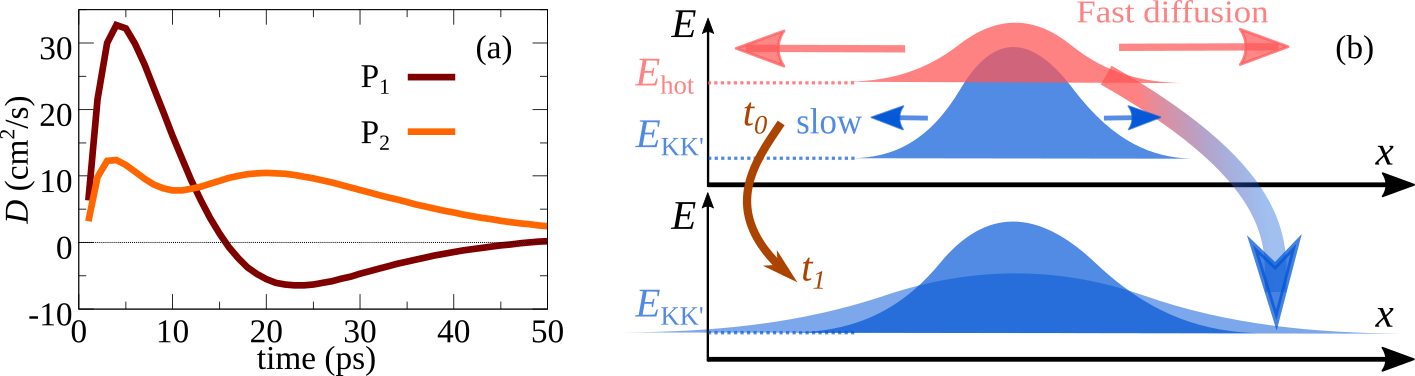}
		\caption{(a) Effective diffusion coefficient at phonon-sideband energies P$_1$ and P$_2$. (b) Sketch showing how  hot exciton thermalization produces a bump in the diffusion coefficient of P$_2$.
		}
	\label{figS1}
\end{figure}

\begin{figure}[b!]
	\centering
			\includegraphics[width=16 cm]{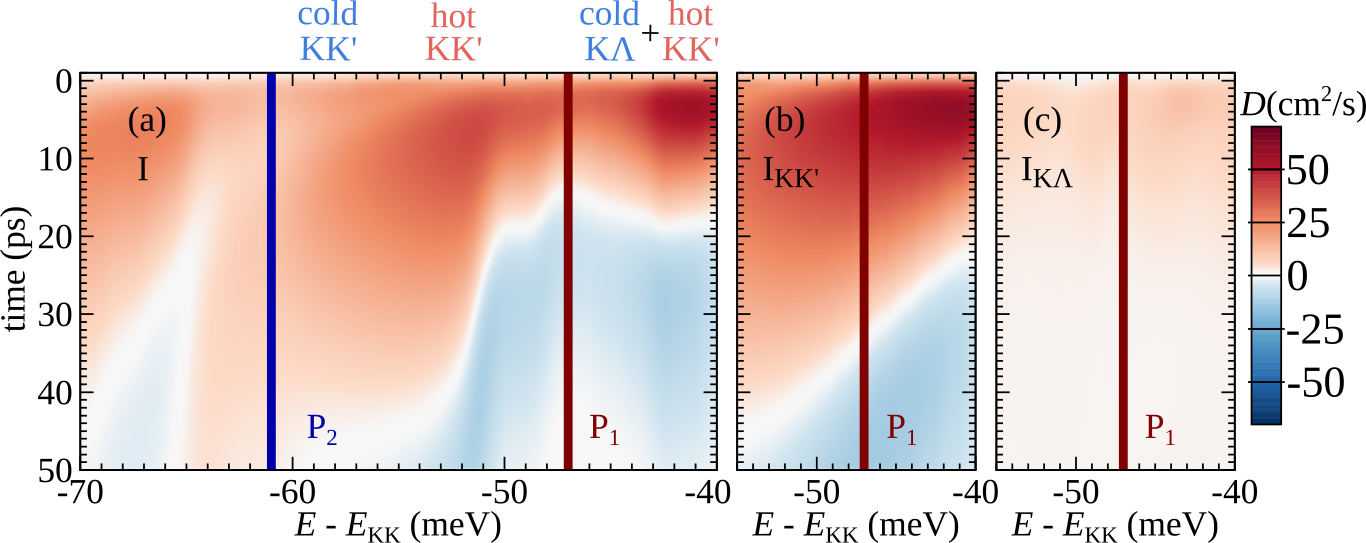}
		\caption{Energy-resolved effective diffusion coefficient  of 
		(a) $I(E,\mathbf{r},t)$ and (b)-(c) focusing on the energy around P$_1$ and separating the contributions from  $I_{KK^\prime}(E,\mathbf{r},t)$ and $I_{K\Lambda}(E,\mathbf{r},t)$, respectively.
		}
	\label{figS2}
\end{figure}

In Fig. S\ref{figS1}(a) we show the  evolution of the effective diffusion coefficients $D_{\text{P}_1}$ and $D_{\text{P}_2}$ at the energies  of the phonon sidebands P$_1$ and P$_2$, cf.  Fig. 2 in the main manuscript. The quick spatial broadening of P$_1$ as shown in Fig. 3(a) of the main manuscript is directly reflected in a steep increase of the diffusion coefficient $D_{\text{P}_1}$ reaching values as high as 35 cm$^2$/s. The steep increase is followed by  a quick decrease, showing also  transient negative diffusion coefficients \cite{Rosati20}. The origin of the high diffusion values can be traced back to  hot excitons in the KK$^\prime$ valley (rather than K$\Lambda$ excitons, cf. discussion on Fig. S\ref{figS2}). Higher effective diffusion coefficients are often followed by negative values as a consequence e.g. of inter- and intravalley exciton thermalization \cite{Rosati20}. 

In contrast, the maximum value  reached by $D_{\text{P}_2}$ is approximately three times smaller, since at the energy $E\equiv$P$_2$ there are no PL contributions from hot excitons. Nevertheless, the latter can still affect the evolution of  $D_{\text{P}_2}$ in an indirect way, as reflected by the bump at approximately 20 ps. The origin of the bump is the thermalization of hot KK$^\prime$ excitons, cf. Fig. S\ref{figS1}(b). Soon after the optical excitation (top panel), hot and cold excitons in the same valley (red and blue area, respectively) have the same spatial width (provided by the optical excitation). However, the hot excitons diffuse faster and hence   show a larger spatial broadening than cold excitons (bottom panel). Once they thermalize into the valley minimum via intravalley scattering, the spatial distribution of cold  excitons suddenly becomes larger resulting in  the bump observed in Fig. S\ref{figS1}(a).
Note that in case of a faster thermalization (e.g. at higher temperatures) 
or slower diffusion (e.g. for broader initial spatial distributions, cf. Fig. S\ref{figS4}), 
we do not see such a bump, since  cold and hot excitons have similar spatial distributions when the latter thermalize.

Fig. S\ref{figS2}(a) shows the theoretically predicted effective diffusion coefficients as a function of energy and time. The diffusion at P$_2$ is dominated by cold KK$^\prime$ excitons located at the valley minimum, while at higher energies hot excitons take over resulting in much  higher diffusion coefficients (dark red regions).
At energies  around  P$_1$, both cold K$\Lambda$ excitons populating the valley minimum as well as hot KK$^\prime$ excitons contribute to the diffusion. 
To separate the two contributions, in Figs. S\ref{figS2}(b-c) we consider only  $I_{\text{KK}^\prime}(E,\mathbf{r},t)$ and $I_{\text{K}\Lambda}(E,\mathbf{r},t)$, respectively. 
We observe very clearly that the initial fast diffusion (dark red regions) clearly stems from $I_{\text{KK}^\prime}$, i.e. hot KK$^\prime$ excitons located approximately at the energy of E$_{\text{K}\Lambda}$.  The signal stemming  from cold K$\Lambda$ excitons shows approximately one order of magnitude smaller diffusion coefficients, cf. Fig. S\ref{figS2}(c). 
Furthermore, the photoluminescence $I_{\text{KK}^\prime}(\text{P}_1,\mathbf{r},t)$ shows a smoother and slower transition to negative diffusion coefficient values (Fig. S\ref{figS2}(b)) compared to the  total PL  $I(\text{P}_1,\mathbf{r},t)$ (Fig. S\ref{figS2}(a)).
The smooth transition is induced by the intravalley thermalization to lower energies, which is spatially non-uniform, since the occupation at lower energies are narrower in space due to slower effective diffusion, cf. Fig. S\ref{figS2}(b).
In contrast, the sharp decrease to negative values observed at approximately 10-15 ps around P$_1$ in Fig. S\ref{figS2}(a)  is the result of the competition between excitons being localized in different valleys and exhibiting different excess energies. 
In the first 15 ps, the P$_1$ signal  is still dominated by hot KK$^\prime$ excitons, hence the total PL $I(\text{P}_1,\mathbf{r},t)$ shows a fast broadening (Fig. \ref{figS2}(a)) in accordance to the hot excitons in KK$^\prime$ (Fig. S\ref{figS2}(b)) and in contrast to the slow propagation of K$\Lambda$ (Fig. S\ref{figS2}(c)).
Once  hot KK$^\prime$ excitons have cooled to energies below $E_{\text{K}\Lambda}$ via intravalley scattering occurring within the first 10-15 ps at 20K, they are not able to contribute to P$_1$ any longer. As a result, the latter  becomes  dominated by cold
K$\Lambda$ excitons, which propagate slower in space. 
Thus, the relaxation of hot KK$^\prime$ excitons to energies below $E_{\text{K}\Lambda}$ results in a sharp decrease of the spatial width of excitons giving rise to the P$_1$ sideband (Fig. S\ref{figS2}(a)).

\section{Dependence on initial spatial broadening}\label{sec:Space}

\begin{figure}[b!]
	\centering
			\includegraphics[width=16 cm]{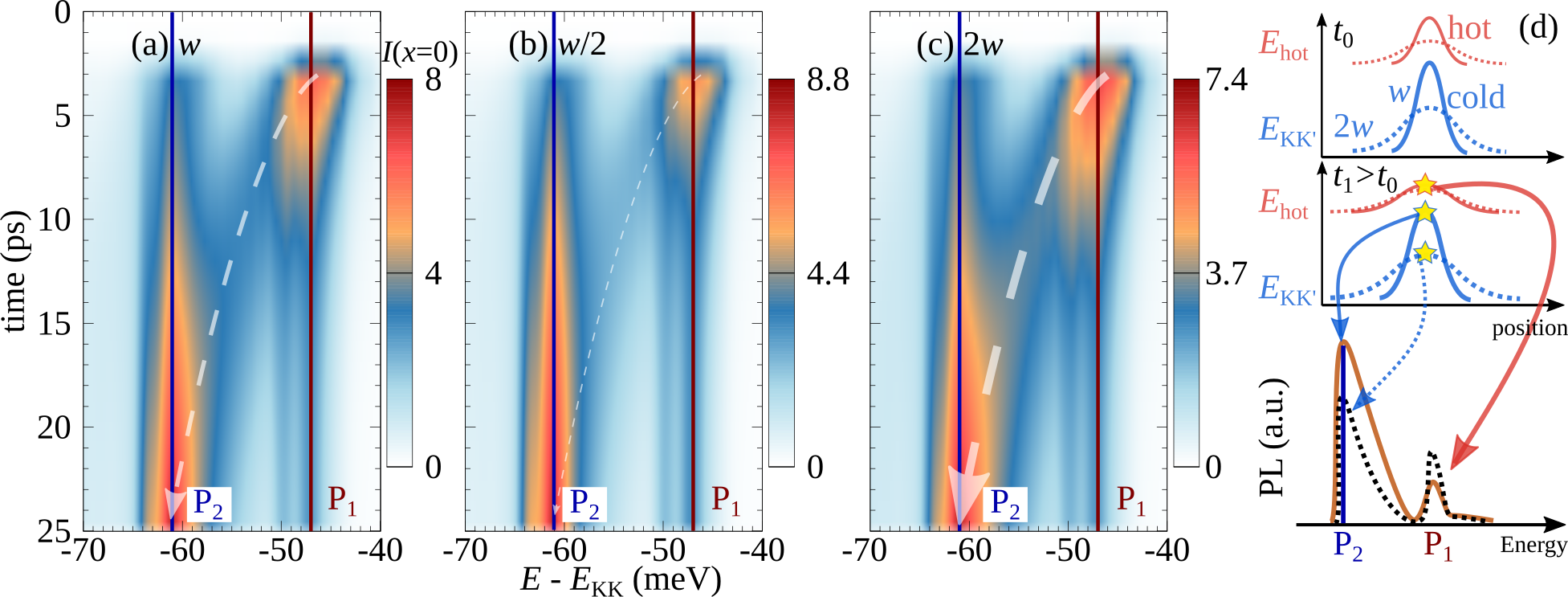}
		\caption{Energy- and  time-resolved photoluminescence $I_{w,0}(E,t)$ coming from the center of an optical excitation $x=0$ with spatial width (a) $w$ (as in the main text), (b) $w/2$ and (c) $2w$. The PL has been normalized to $I(E_{\text{KK}},x=0,t)$ (with the colorbar maximum adjusted to the value of P$_2$ at 25 ps). (d) Sketch showing how a narrower spatial distribution (solid lines) provides a faster diffusion especially for hot excitons (red lines), while for a broader spatial distribution (dotted lines) hot and cold excitons show a similar broadening (upper panel). This results in a lower PL intensity for a narrower spatial distribution evaluated  at $x=0$ at  P$_1$, which is initially  ($t\lessapprox$10 ps) dominated by hot excitons  (bottom panel).
		}
	\label{figS4}
\end{figure}

The faster diffusion of hot excitons results in spatially dependent occupations, i.e. hot excitons are  more present in the spatial tails than at the center of the optically excited spatial distribution, cf. Fig \ref{figS1}(b). 
To study this we calculate spectrally-resolved PL at the center of an excitation pulse ($x=0$) with a varying spatial width $w$.
%\commentAC{Here, it would be good to explicitly clarify that FHWM $\Delta$ refers to the spatial size, to avoid any misunderstanding especially since this sentence has ``spectrally-resolved'' PL in it. Also, since the width $w$ parameter is otherwise used throughout the paper and the manuscript, it would be useful to state the ratio between the two.}
In Fig. S\ref{figS4}(a) we show  $I_{\Delta,0}(E,t)$ with  $w\approx0.5\mu$m as in the main manuscript. We find that  the PL is initially ($t\lessapprox$10 ps) dominated by  the P$_1$ phonon sideband (red line), while at later stages the P$_2$ phonon sideband (blue line) becomes dominant. In the transient regime we observe a thermalization process of hot excitons initially formed approx. 15 meV above $E_{\text{KK}^\prime}$ [see Figs. 2(a-c) of main manuscript]. As they relax towards the minimum of KK$^\prime$ excitons, the  
center of the corresponding PL shifts to the red (dashed arrow) resulting also in a transfer of the
optical weight from P$_1$ to P$_2$.

Figures S\ref{figS4}(b) and (c) illustrate the same study however now decreasing (increasing) the spatial width $w$ of the initial spatial exciton distribution to $w/2$ (2$w$). In the case of a narrower distribution we observe a much 
weaker transient ($t\lessapprox$10 ps)
signal stemming from P$_1$ and a less pronounced relaxation process of hot exctions from P$_1$ to P$_{K^\prime}$ (cf. the thin dashed arrows). The reason for this behaviour lies in the interplay of the thermalization in energy  and the diffusion in space.
The narrower the exciton occupation, the larger  its gradient and the faster is the diffusion, cf. Fig. \ref{figS4}(d). This applies in particular to faster-diffusing hot excitons, while cold excitons with a small excess energy are less affected.
Therefore, a smaller initial spatial width  leads to a faster diffusion of hot excitons away from the center at $x=0$, where their population becomes depleted. As a consequence, the phonon-assisted transient signal 
from hot excitons at e.g. 
P$_1$ is weaker compared to a larger initial $w$ (solid vs dashed line in the bottom panel of Fig. \ref{figS4}(d)).
The situation is reversed for an initially broader spatial distribution: Here the diffusion is weaker, hence spatial distributions of hot and cold excitons are similar (dotted lines in Figs. S\ref{figS4}(d)). This results in a larger presence of hot excitons in $x=0$, resulting in a larger transient signal at P$_1$, cf. Figs. S\ref{figS4}(a,c).

Note that while the diffusion leads to a depletion of hot excitons from the center of the distribution  $x=0$, the opposite takes place in the spatial tail of the distribution, as hot excitons diffuse  there from  $x=0$.  
In summary,  hot excitons move away from the center of the optical excitation. This is particular true for narrower initial distributions, leading to a weaker corresponding spectrally-resolved emission from the center of the optical excitation.

\section{Low-temperature effective diffusion in MoSe$_2$}\label{sec:Space}

\begin{figure}[t!]
	\centering
			\includegraphics[width=16 cm]{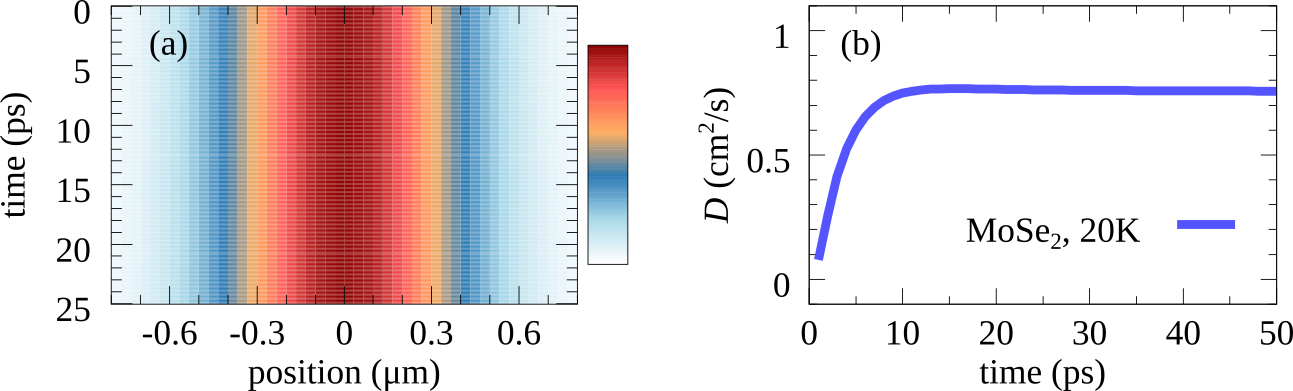}
		\caption{(a) Spectrally-integrated PL in MoSe$_2$ at 20 K  and (b) associated effective diffusion coefficient $D$ displaying no onset of fast effective diffusion due to the absence of hot excitons. 
		}
	\label{figSMoSe}
\end{figure} 

While the speed-up of the effective diffusion in WSe$_2$ has been explained by hot excitons formed in the energetically lower-lying dark valleys, in Fig.   S\ref{figS4} we show the behaviour in MoSe$_2$. Here KK is the energetically lowest valley, hence no dark hot excitons are initially formed. As a result, no initial speed-up of the energy integrated PL is observed in the first picoseconds (Fig.   S\ref{figS4}(a)), as revealed also by a quantitative analysis of the effective diffusion coefficient, Fig.   S\ref{figS4}(b). Here  the maximum value coincides with the stationary one, in clear contrast with the case of WSe$_2$ where the two differed by an order of magnitude. The absence of initial fast effective diffusion in MoSe$_2$ shows once again that such a speed-up is as obtainable only in presence of dark valleys, where transient hot excitons are formed.

\section{Experimental photoluminescence spectra}\label{sec:ExpSpectra}

\begin{figure}[t!]
	\centering
			\includegraphics[width=8.5 cm]{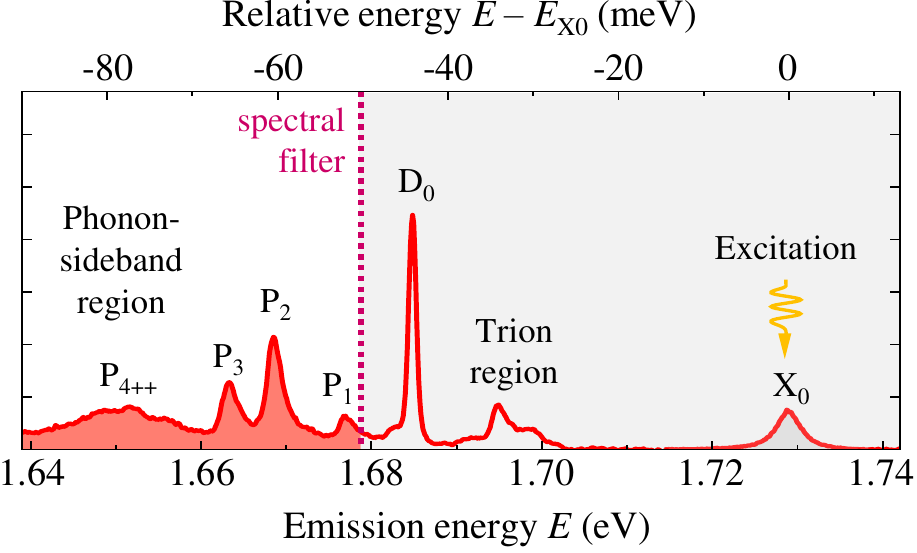}
		\caption{Time-integrated PL spectrum of a hBN-encapsulated WSe$_2$ monolayer at T = 5\,K. 
		Direct exciton recombination of bright and out-of-plane polarized spin-dark KK excitons is labeled by X$_0$ and D$_0$, respectively.
		Phonon sideband emission features are labeled by P$_1$ $-$ P$_{4++}$. 
		For the diffusion measurements the detected light was spectrally cut-off by a long-pass filter, blocking the PL indicated by the gray area.
		}
	\label{figS5}
\end{figure}

In the main manuscript, the theoretically predicted fast diffusion of hot excitons is experimentally demonstrated  by time- and spatially-resolved PL measurements on a hBN-encapsulated monolayer WSe$_2$.  
For an overview, a typical time-integrated, unpolarized luminescence spectrum of the studied sample at T = 5\,K is shown in Fig. S\ref{figS5}. 
Below the bright X$_0$ transition at almost 1.73\,eV we observe a series of characteristic emission peaks. 
These include weak PL from negatively charged trion doublet about 30\,meV below X$_0$ including an additional peak in-between, as well as the direct recombination from spin-dark excitons polarized in the out-of-plane direction and labeled as $D_0$.
Below $D_0$ are several features originating from phonon-sideband emission of dark excitons in monolayer WSe$_2$.
The most prominent resonance in this regime, about $60$ meV below the bright exciton and labeled by P$_2$, represents phonon-assisted exciton recombination of momentum-indirect KK$^\prime$ excitons\cite{Liu19, He20}. 
P$_3$ and P$_{4++}$ were recently argued to stem from the phonon replicas of spin-dark KK \cite{Liu19} and momentum-indirect KK$^\prime$ excitons under optical phonon emission, respectively \cite{Brem20, Rosati20b}. 
Finally, at energies around P$_1$ both hot KK$^\prime$ and cold K$\Lambda$ excitons should contribute to the emission \cite{Rosati20b}.

For diffusion measurements the monolayer was resonantly excited at X$_0$ using a $100$\,fs Ti:sapphire laser.
The PL signal was detected with a streak camera, providing both spatial and temporal resolution.
Here, we focused on the diffusion of momentum-dark excitons by monitoring their phonon-assisted emission. 
To cut-off direct exciton recombination of the D$_0$, trion, and X$_0$ features, we used a spectral filter to detect only the phonon-sideband region, as indicated in Fig. S\ref{figS5}

\section{Transient diffusion coefficient estimated from time-resolved spectra}\label{sec:DiffusionSpectral}

Here, we illustrate the estimation of transient diffusion coefficients based on the time-dependent measurements of emission energies from the PL spectra.
In the semi-classical description, that should be applicable at the studied temperature of 5\,K, the diffusion coefficient $D$ has the form: 
\begin{equation}
D=\frac{\langle E_{kin}\rangle \tau_s}{M_x}.
\label{semiclass}
\end{equation}
It is proportional to the mean kinetic energy $\langle E_{kin} \rangle$ of the excitons, momentum scattering time $\tau_s$, and the exciton total mass M$_x$. 
For an equilibrated exciton distribution the kinetic energy is given by the lattice temperature, i.e. $\langle E_{kin} \rangle = k_BT$.
At non-equilibrium conditions, however, it can be approximated by adding an average excess energy $\Delta E$.
The latter can be estimated from the spectral analysis, since phonon-assisted processes allow for recombination of excitons with essentially arbitrary kinetic energies and momenta.
Consequently, $\Delta E$ should essentially correspond to the relative shift of the phonon sideband emission energy with respect to the equilibrium value\,\cite{Rosati20b}.

 \begin{figure}[h!]
	\centering
			\includegraphics[width=15 cm]{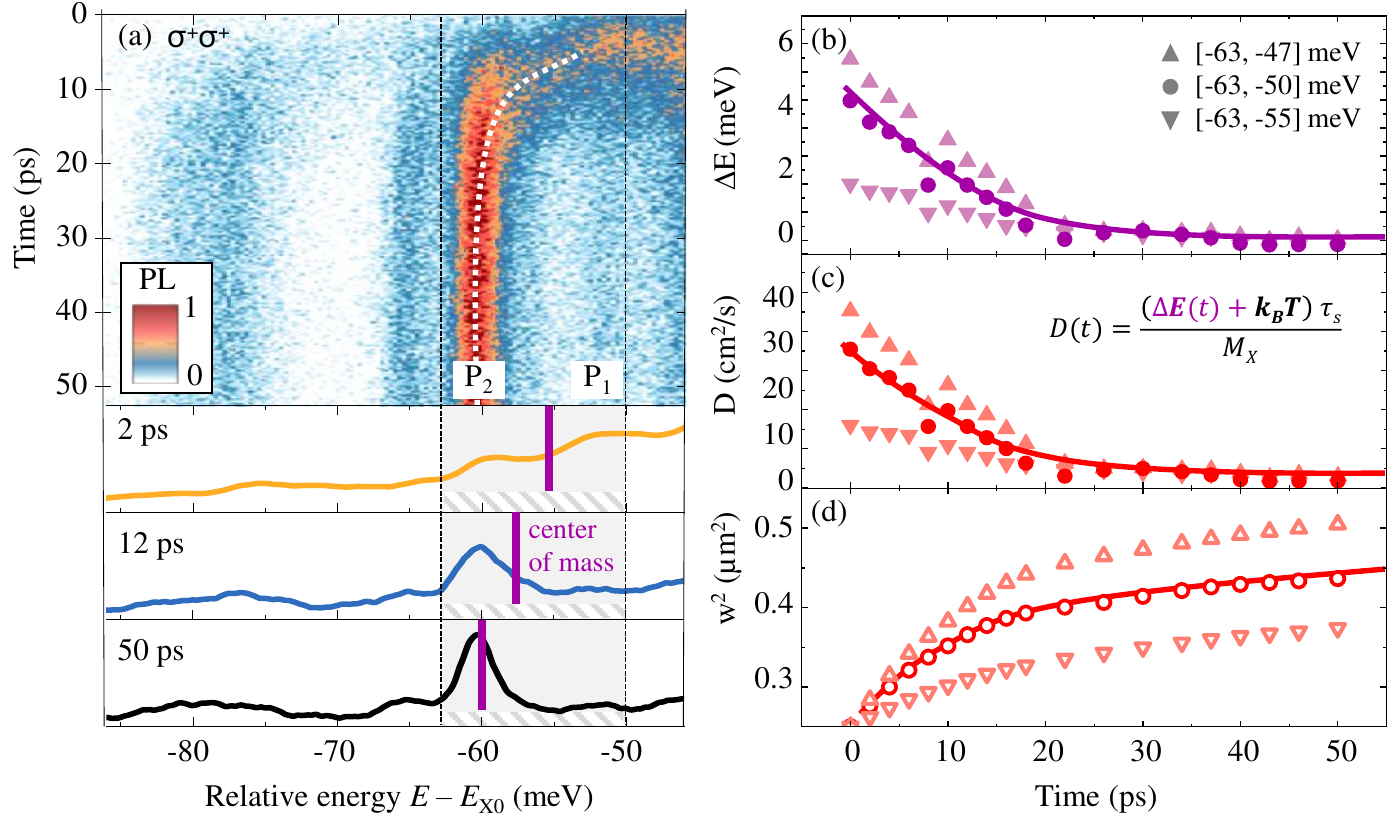}
		\caption{Transient diffusion coefficients from spectral energy shifts.
(a) Experimental time- and energy-resolved PL spectra of a hBN-encapsulated WSe$_2$ monolayer at T $=5$ K after circularly polarized, resonant excitation of X$_0$.
Data is shown for co-polarized detection in the energy range where only phonon side bands contribute to the PL (taken from Ref. \cite{Rosati20b}). 
The dashed white line schematically illustrates the average energy shift $\Delta$E with time. 
Lower panels show PL spectra at selected times. 
Purple lines indicate center of gravity within the selected energy range between $-63$ and $-47$ meV, as indicated by the gray area.
(b) Extracted energy shifts $\Delta$E(t) from (a) for several energy intervals with different values of the high-energy cut-off. 
(c) Corresponding transient diffusion coefficient $D$ obtained via the semi-classical expression in Eq.\,\ref{semiclass} using parameters $\tau_s = 2.8$\,ps and $M_x = 0.75 m_0$.
(d) Resulting time-dependent broadening of the spatial profile for an initial spot width of $w\approx0.5$ $\mu$m.
   }
	\label{figS6}
\end{figure} 

For the purpose of this analysis we consider time-resolved PL spectra of the phonon sidebands, obtained for circularly polarized resonant excitation and co-polarized detection.
The data is reported in a previous study\,\cite{Rosati20b} and reproduced in Fig. S\ref{figS6}(a).
The emission in the spectral range between -45 and -65 meV with respect to X$_0$ largely stems from the phonon-assisted recombination of the momentum-dark KK$^\prime$ excitons (c.f., P$_1$ and P$_2$ features in Fig. S\ref{figS5}).
For an exemplary illustration we neglect additional contributions from K$\Lambda$ excitons with small excess energies that do not significantly influence fast initial diffusion.
As discussed in the main manuscript and Ref.\,\cite{Rosati20b}, the 
transient energy shift of the emission is a direct consequence of initially overheated exciton distribution of KK$^\prime$ states that subsequently equilibrate and cool down over time.  
The energy of the shift of PL signal in this range, relative to the steady-state emission of the $P_2$ phonon sideband at -60\,meV, thus roughly corresponds to the average excess energy $\Delta E$ of the KK$^\prime$ excitons.

From the spectra, we determine the $\Delta E$ shift by evaluating the center of gravity, as illustrated in the exemplary data in the bottom panel of Fig. S\ref{figS6} (a).
Here, we use the high energy cut-off of -50\,meV corresponding to the conditions in the diffusion measurement. 
A constant offset is subtracted prior to the analysis and the resulting mean energy is marked by purple lines.
The resulting the mean energy shift $\Delta E$ is presented as a function of time Fig. S\ref{figS6} (b).
For the reference, we also include smaller and larger values for the high-energy cut-off.
Diffusion coefficients presented in Fig. S\ref{figS6} (c) are then obtained via Eq.\,\eqref{semiclass} by setting $\langle E_{kin} \rangle = \Delta E + k_BT$ using the $\Delta E$ values from Fig. S\ref{figS6} (b).
We further set the momentum-scattering time $\tau_s$ to $\hbar/(0.235$\,meV$) = 2.8$\,ps, corresponding to the value for the exciton scattering with linear acoustic phonons, extracted from temperature-dependent linewidth broadening coefficient of 47\,$\mu$eV/K.
The exciton total mass of 0.75 m$_0$ is taken from the sum of the electron and hole masses in monolayer WSe$_2$ \cite{Kormanyos15} and the lattice temperature is set to the heat-sink temperature of T = 5\,K.
Fig. S\ref{figS6}(d) shows the resulting transient broadening of the spatial PL width assuming an initial Gaussian distribution with $w_0=0.5$ $\mu$m and $w^2=w_0^2+4\int \text{D} dt$.

From this analysis, we indeed confirm the expectation of an initially rapid expansion of a hot exciton cloud with diffusion coefficients on the order of 20 to 40 cm$^2$/s at early times.
Moreover, the equilibration time-scale of about 10 to 15\,ps is very similar to both theoretical predictions and the decay of transient diffusivity in spatially-resolved measurements.
As discussed in the main manuscript, such close quantitative agreement with direct measurements of the diffusion provides further support for the overall consistent interpretation of our observations.

\bibliography{rosatiBib_short,references}